# Evolution of Electrical Resistivity, Thermal Conductivity, and Temperature of a solid under the action of Intense Ultrashort Laser pulse

Arvinder S. Sandhu, A. K. Dharmadhikari and G. Ravindra Kumar*

*Tata Institute of Fundamental Research, 1 Homi Bhabha Road, Colaba, Mumbai, INDIA 400005*

**The dynamical properties of Cu in a regime relevant to femtosecond micro machining are obtained on picosecond time scales using pump-probe reflectivity study for 100fs, $10^{15}$ W $cm^{-2}$ laser pulses. The electrical resistivity is obtained by solving Helmoltz equations. The dissipation mechanisms and scaling laws are obtained in high and low temperature limits. The 'resistivity saturation' effect in an unexplored regime intermediate to hot plasma and cold solid is studied in detail. The temperature evolution and thermal conductivity is obtained in the temporal range 0 to 30ps after the interaction of laser pulse with solid Cu.**

* e-mail: grk@tifr.res.in

The availability of femtosecond lasers has sparked a huge amount of work in both fundamental and applied physics apart from numerous applications in other fields as well. There has been a great deal of interest in understanding the evolution of a solid impulsively heated by a sub-picosecond laser pulse thereby creating highly non-equilibrium conditions[1,2]. Substantial understanding of electronic properties of solids in non-equilibrium and short time scales regimes has come from relatively low intensity (~$10^{12}$ $Wcm^{-2}$) experiments[1-5]. At sufficiently high intensity, phenomena like ultrafast phase transitions, thermal and non-thermal structural transformations and melting dynamics have been attracting significant attention[6-8]. The dynamics of solid evolution for the laser pulse intensities ≥$10^{14}$ $Wcm^{-2}$ has immediate importance in view of the increasing application of femtosecond laser pulses for precision micro machining and material processing[9-12] and to our knowledge no work has been reported on time resolved evolution of solid in this context.

As a general framework, it has been established that when a short laser pulse hits a solid, the electrons by virtue of their lower heat capacity than the lattice ( at least by one or two orders of magnitude) are elevated to 1-2 orders higher temperature than the lattice[1,2]. Within the electron subsystem the energy exchange timescales are of the order of $10^{14}$ $s^{-1}$ for $T_e > 1$eV, thus electrons are in local thermodynamic equilibrium and can be characterized by a temperature value. Strongly coupled ions can also be similarly described by a temperature value. Thus the situation can be approximately modeled as a two-temperature non-equilibrium system. The understanding of laser energy absorption under such situations is crucial as it can help determine electron and thermal transport, equation of state, structural properties and phase transitions. Time resolved reflectivity is thus a powerful tool to extract dynamical information about non-equilibrium system. A very interesting example of the application of time resolved reflectivity[13] is demonstration and measurement properties of the fluid phase of carbon, which exists only in the very high temperature and pressure regime.

Here, we use time resolved femtosecond pump-probe reflectivity to determine the dc-electrical resistivity of Copper metal. We can classify our study into three regimes, namely relatively cold solid, hot plasma and the intermediate case. The intermediate transition regime exhibits the phenomenon of 'resistivity saturation', where the electron mean free path becomes of the order of inter-particle spacing. The resistivity saturation was invoked by Milchberg et. al.[14] in an single pulse experiment on solid Aluminum. Here we do time resolved study of this regime and also establish the mechanisms of energy damping and temperature scaling of various processes. In the intermediate regime one cannot use simple weakly coupled plasma description and the collision frequency (which determines energy dissipation) has to be determined rigorously taking care of strong coupling between electrons as well as between ions[15]. To our knowledge, there exists no such general formalism to encompass the transition from hot plasma to cold solid over picosecond timescales. It is known that the collision frequency ($\nu$) scales as $T_e^{-3/2}$ for hot plasma limit, where '$T_e$' is the electron temperature. For low temperature equilibrium conditions $\nu$ is expected to arise from electron-phonon collisions mainly and $\nu \propto T_i$, where '$T_i$'is ion temperature. However, when electron temperature is more than Fermi temperture and much higher than ion temperature, the electron-electron scattering which goes as $T_e^2$ is generally assumed to be dominating over electron-phonon contribution.[16-18]

The behavior of metal under the action of laser pulse i.e. absorption processes, and the evolution of electron and ion temperature, can be modeled by the following set of equations[2,17]:

$$C_e(T_e)\frac{\partial T_e}{\partial t} = k\nabla^2 T_e - g(T_e - T_i) + A(r,t) \tag{1}$$

$$C_i\frac{\partial T_i}{\partial t} = g(T_e - T_i) \tag{2}$$

$$\frac{d^2E}{dx^2} + k^2(n^2 - \sin^2 q)E = 0 \tag{3}$$

$$\frac{d^2B}{dx^2} + k^2(n^2 - \sin^2 q)B - i\frac{4p}{ckn^2}\frac{ds}{dx}\frac{dB}{dx} = 0 \tag{4}$$

Equations (1) and (2) are coupled equations for the electron and ion temperatures respectively. Here $C_e$ $(C_i)$ is electron (ion) heat capacity, '$g$' is electron phonon coupling constant, '$k$' is thermal conductivity and 'A' represents the laser energy absorption. The equation (3) and equation (4) are Helmoltz equations for propagation of s and p-polarized laser pulse respectively in a finite scale length density gradient inside the metal. The refractive index '*n*' for Drude (intraband) absorption is of the form, $n^2 = 1 + i4ps / w$, 'ω' is the laser frequency and 'σ' is the conductivity given as $s = \frac{1}{4p}(n + iw)[w_p^2 /(w^2 + n^2)]$, where $w_p = 4pn_e e^2/m_e$, the plasma frequency.

For our experiment we obtain a high intensity (3 x $10^{15}$ W/cm$^2$, 100fs, 806nm) pulses from a custom built chirped pulse amplification Ti:S laser which is described in detail elsewhere[19]. A small part of laser is frequency doubled in a thin BBO crystal to 403nm and used as probe. The laser pulse contrast was $10^{-5}$ at 1ps for the fundamental and is expected to be $10^{-10}$ for the second harmonic probe. The targets used in these experiments were polished Cu discs with flatness better than λ/5. The delay between pump and probe is changed in small steps using a precision translation stage. Two similar photodiodes (Hamamatsu) are used to measure the reflected and input laser energy variations. In the data presented here we monitored the temporal behavior of probe reflectivity from –5ps to +30 ps time delay. The reflectivity of probe was found to remain specular within our range of interest. Figure 1 shows the time resolved reflectivity of a weak ($10^{12}$ W cm$^{-2}$) p-polarized probe at near normal incidence from a copper target excited by 3 x $10^{15}$ W cm$^{-2}$, p-polarized pump pulse incident at 50$^o$. Copper is interesting system as it allows us to distinguish processes of pump-induced fast electron heating (sharp transient in reflectivity from 0.5 to 0.8 near zero delay) and slower electron cooling afterwards.

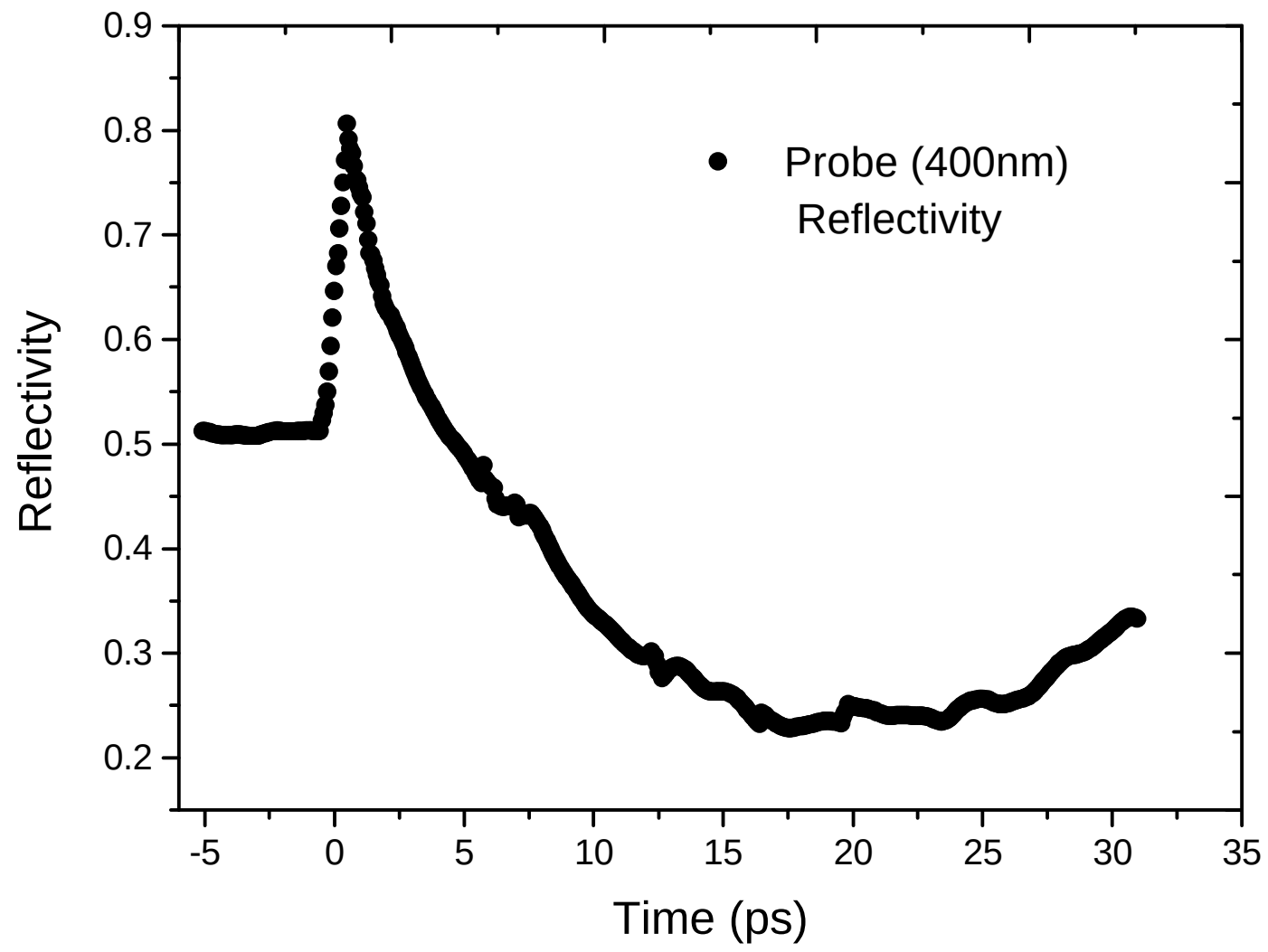

Fig1: Time resolved reflectivity of probe (λ=403nm, I=$10^{12}$ Wcm$^{-2}$) from Copper irradiated by 3 x $10^{15}$ Wcm$^{-2}$, 806nm pump pulse incident at 50$^o$.

This is because of strong inter-band absorption near 400nm as seen on the negative delay side, thus giving low ‘cold’ reflectivity. This distinction is not possible in a simple metal (e.g. Al) as the ‘cold’ reflectivity is very high and thus a fast rise in reflectivity transient is does not appear prominent. At the peak value of reflectivity the Cu is in hot plasma state and the reflectivity is solely determined by Drude absorption[17].

For quantitative understanding we numerically solve Helmoltz equation following Milchberg et. al.[20], for the p-polarized probe propagating inside hot evolving plasma. The density profile is assumed to be of the form $n_e = n_{solid}(1 - x/L)^2$, where $n_{solid}$ = 8.47 x $10^{22}$ $cm^{-3}$. The plasma scale length *L(t)* is estimated from Doppler shift measurements of probe and is also consistent with electron temperature profiles discussed later. Then for each reflectivity value, we solve (equation (4) above) for the value of collision frequency parameter ν needed to get corresponding reflectivity at that instant of time (Fig 3). However, the more important and commonly used parameter is dc-resistivity, which is related to collision frequency as $r_{dc} = nn_e / me^2$ as can be seen from the expression for σ. For the evolution of temperature we then solve coupled diffusion equations for electron and ion temperatures. The thermal conductivity κ is taken to be of the form[21] $k = k_o(T_e/n)$, where we have $\kappa_o$ = 7.5 x $10^{13}$, obtained as in Ref. 22. The electron specific heat $C_e$ goes as 96.6 $T_e$ $Jm^{-3}K^{-1}$ at low temperatures[5] and as $1.5k_Bn_e$ at high temperatures[17]. We use harmonic interpolation for intermediate temperatures. The ion

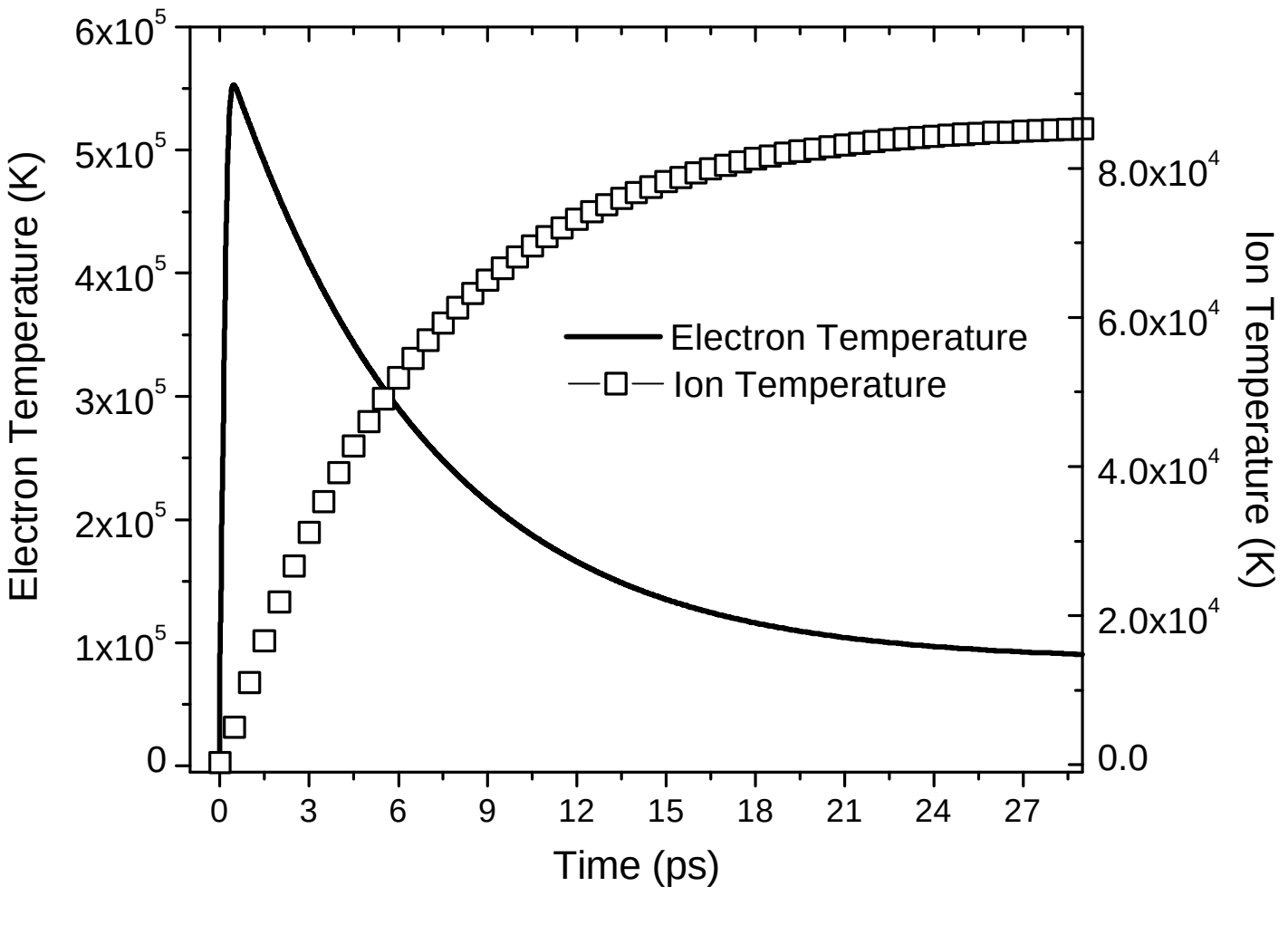

Fig2: The electron (line) and ion (open squares) temperature evolution as a function of time delay.

specific heat $C_i = 3\, k_B n_i$ and coupling constant '$g$' is taken as $10^{17}$ $Wm^{-3}K^{-1}$ from Ref. 5. The laser energy absorption is incorporated assuming a gaussian intensity profile of laser pulse with FWHM of 100fs. The resulting electron and ion temperature profiles at the solid density are as shown in figure 2. The electron temperature sharply increases to about 50 eV in less than a picosecond while ions are still two orders of magnitude lower in temperature, thus resulting in highly non-equilibrium situation. The dynamics after pump pulse is determined by electron lattice coupling and electron thermal conductivity. As seen from figure 2, the electron temperature decreases fast, and substantial equilibration of electron-ion system takes place in 25 ps. It is interesting to note that the ion temperature reaches melting temperature of Cu before the peak of pump pulse is reached. Thus the lattice gets disordered via 'ultrafast thermal melting' in our regime of study.

The dc electrical resistivity and collision frequency discussed above is shown in figure 3 as a function of both electron temperature and time. For our conditions the variation is observed to be between 200 to 500 μΩ-cm. Two limiting cases can be clearly seen within our study. As expected at hot plasma limit where $T_e >> E_F/k_B$, (where $E_F$ is Fermi energy) the collision frequency and hence resistivity goes as $T_e^{-3/2}$. At lower temperature side ($T_e \sim E_F/k_B$), the resistivity varies as $T_e^2$, which shows that the electron-electron scattering which goes as $T_e^2$ dominates over the electron phonon scattering, which varies as $Ti$.

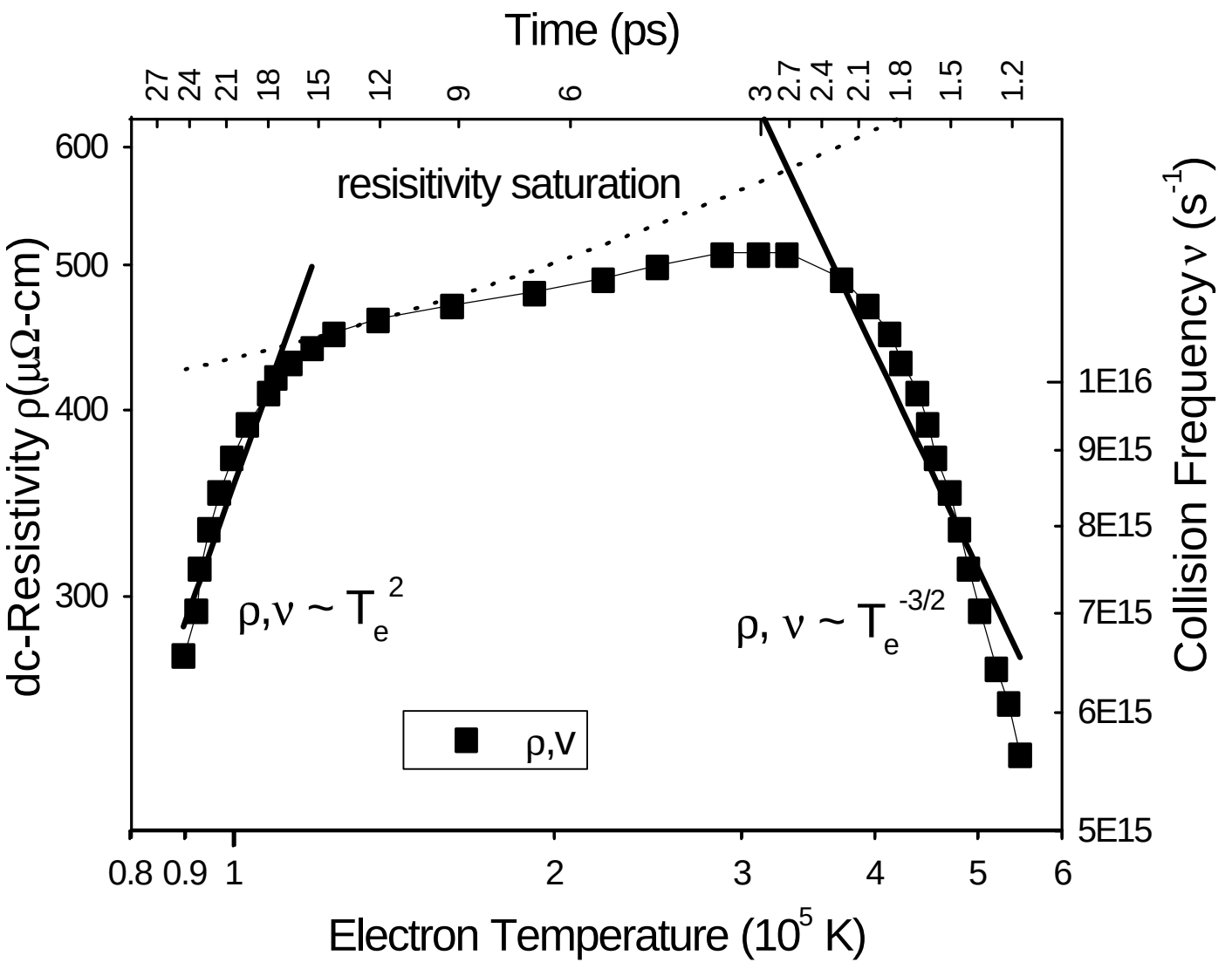

Fig3: The electrical resistivity and collision frequency as a function of time delay (top axis) and electron temperature (bottom axis). The solid lines are for appropriate scaling laws and the dotted curve is the theoretical upper bound on resistivity.

However it is important to note that the system spends most of its time (3 –15 ps) in intermediate temperature regime, thus making it even more important.

The intermediate transition regime shows very little variation and represents saturation of resistivity. The maximum value for resistivity (collision frequency) can be determined by the condition that mean free path $l_e$ between two collisions is greater than the inter-particle distance $r_0$, which can be calculated as $(3/4pn_e)^{1/3}$. This condition can be also written as $n \le V_e / r_0$, where $V_e$ is characteristic electron velocity, which can be in general deduced as, $V_e = \sqrt{V_{Fermi} + k_B T_e / m}$ and $V_{Fermi}$ is the Fermi velocity. The estimated maximum of ν and ρ is plotted in figure 3 as dotted line. The experimental saturation behavior is evidently in good agreement with this calculation.

The thermal conductivity $k = k_o(T_e / n)$ is also deduced as shown in figure 4. The high temperature shows the expected behavior $\sim T_e^{5/2}$. At lower side of temperature scale, it goes as $T_e^{-1}$, again as expected from $k \propto T_e / n$. In order to observe the scaling as $T_e/T_i$ i.e. where electron phonon interaction starts dominating momentum relaxation one has to go to even lower temperatures. However it is not possible to allow unlimited time for further cooling to even lower temperatures in this experiment as the non-specular scattering starts becomes increasingly significant.

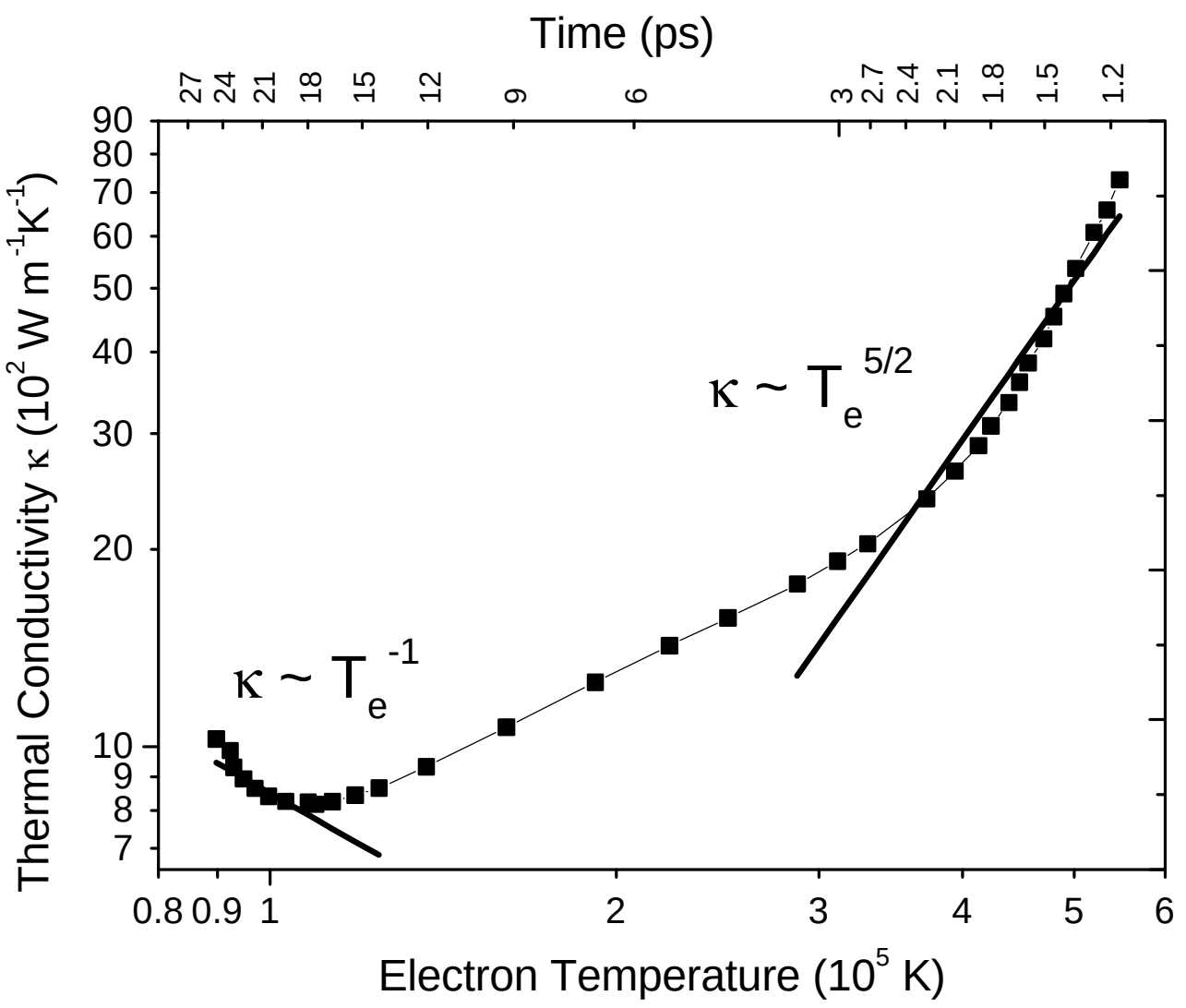

Fig4: Thermal conductivity deduced as a function of time delay and electron temperature. The higher and lower end scaling is shown as solid lines.

In conclusion, from a time resolved pump-probe reflectivity we have deduced electron and ion temperature evolution, the dc-electrical resistivity and thermal conductivity of a solid irradiated by an intense ultrashort laser pulse. No free parameters were used in obtaining the results. At high temperatures collision frequency goes as $T_e^{-3/2}$ as expected for hot plasma. At lower temperatures we establish the importance of electron-electron scattering which goes as $T_e^{-2}$ as the principle energy dissipation mechanism over the electron-phonon scattering. The saturation resistivity value (~450 μΩ-cm) in the intermediate regime agrees well with theoretical estimates. We hope that knowledge of temperature and time dependent evolution of these parameters will be useful for emerging applications involving the use of ultrashort laser interaction with metals.

We thank Sudip Sengupta and P. K. Kaw for discussions and valuable help in numerical codes. We also thank P.P. Rajeev, Vinod Kumarappan, M. Krishnamurthy and D. Mathur for useful suggestions. The laser system used was funded in part by Department of Science and Technology, Govt. of India.